\documentclass[superscriptaddress,showpacs,aps,prd,notitlepage]{revtex4-1}

\usepackage{amsmath,amssymb,color,epsfig, graphicx, epstopdf}

\DeclareGraphicsExtensions{.eps, .pdf,.png,.jpg}

\begin{document}
\title{Exact Relativistic Magnetized haloes around Rotating Disks}

\author{Antonio C. Guti\'errez-Pi\~{n}eres}
\email[e-mail:]{acgutierrez@correo.nucleares.unam.mx}

\affiliation{Facultad de Ciencias B\'asicas,\\
Universidad Tecnol\'ogica de Bol\'ivar, Cartagena 13001, Colombia}

\affiliation{Instituto de Ciencias Nucleares, Universidad Nacional Aut\'onoma de M\'exico,
 \\AP 70543,  M\'exico, DF 04510, M\'exico}

\author{Abra\~{a}o J. S. Capistrano}
\email[e-mail:]{abraao.capistrano@unila.edu.br}
\affiliation{Federal University of  Latin-American Integration, \\ Technological Park of Itaipu,
PO box 2123, Foz do Igua\c{c}u-PR 85867-670, Brazil}

\begin{abstract}
The  exact  relativistic  treatment  of a rotating disk surrounded by a  magnetized material  halo is  presented. The  features of  the halo  and  disk  are described  by  the 
distributional   energy-momentum tensor  of  a  general fluid in  canonical form. All the relevant  quantities  and  the metric and  electromagnetic potentials are  exactly 
determined  by an arbitrary harmonic function only. For instance, the generalized  Kuzmin-disk potential is  used. The particular class of  solutions obtained is  asymptotically 
flat and satisfies  all the  energy  conditions. Moreover, the motion of  a  charged  particle on  the  halo  is  described.  As  far as we know,  this  is  the  first 
relativistic  model describing analytically the  magnetized  halo of a rotating  disk.  
\end{abstract}

\maketitle
\section{Introductory remarks}
In the observational context, many ambiguities still exist  about the main constituents,  geometry and  dynamics (thermodynamics) of the galactic  disk-haloes. However,  there  are 
several  different observations  which  probe the  galactic and surrounding galactic magnetic  field. A current  revision  of  the status of our knowledge about the magnetic fields 
in our Milky Way and in nearby star-forming galaxies is summarized in \cite{beck2015magnetic}. Additionally, a study of  the   disk and halo rotation are reported  in 
\cite{de2014planck}, whereas the possibility of  magnetic fields can be generated in the outskirts of disks is  studied  in \cite{mikhailov2014magnetic}. Solutions for the Einstein 
and Einstein-Maxwell Field Equations  which are consistently applicable to the context of astrophysics remains a topical problem. Nevertheless, the effects of magnetic fields on 
the physical processes in  galaxies and their disk-halo interaction have been scarcely considered in the past. Similarly, the  relevance of relativistic models of  disks around 
black holes in a  magnetic  field is  discussed in \cite{gutierrez2014exact}.

The presence of  the  electric  field on the dark matter halo models has been considered  in \cite{Chakraborty:2014paa},  whereas  the  presence of  electromagnetic  field in the 
halo-disk  system has  been  studied in \cite{PhysRevD.87.044010,  gutierrez2013variational} in which the gravitational sources are statics. In 
\cite{PhysRevD.87.044010}  we provided an detailed  overview  of the research  in  the relativistic  disks,  accordingly we shall not repeat them here. 

In this paper  we   considered  the conventional treatment of galaxies modelled as a  stationary  thin disk and,  correspondingly, we  associate the  halo  with the region  
surrounding the disk. We present  the  conformastationary version of the static thin disk-halo systems studied in  \cite{PhysRevD.87.044010}. In  addition, we study  the features 
of the  principal quantities characterizing the  dynamic  of  the magnetized  haloes corresponding to  the  disks  presented  in \cite{2015GReGr.47.54G}.  Therefore, we take  the  
definition in Ref. \cite{stephani2003exact}  as  standard,  following the  original  terminology  by  Synge \cite{synge1960relativity}: conformastationary  are  those stationary 
spacetimes with a  conformally flat space of  orbits. 

Accordingly, we  show that the  rotating  disk-haloes  with isotropic pressure, stress tensor and  heat flow generalize  the  static disk-haloes  obtained in  
\cite{PhysRevD.87.044010}.  Our results  are  compatible with those presented in \cite{Chakraborty:2014paa} on possible features of galactic  halo. Moreover, the description of 
the motion of charged  particles on disk is deduced  and  is in agreement with the  results of the similar  analysis  discussed  in \cite{garcia2014exact}.  As far as we know, 
this  is the  first relativistic  model describing analytically the relativistic magnetized  halo of a rotating disk.   

The  paper is  organized  as follows. In  Section \ref{sec:formalism},  the distributional Einstein-Maxwell equations  for  haloes  surrounding thin disks are  obtained. In 
Section \ref{sec:EMT-DH} we  obtain  expressions, in terms of  an arbitrary harmonic  function,  for  the most important quantities characterizing the dynamic  of the disk and  
halo. In Section \ref{sec:examples}  we  first  calculate  quantities  for  an harmonic  function  described  by  the generalized Kuzmin-disk potential. Then, we analyze  the  
obtained results and  calculate the  constants of motion of  the   disk. Moreover, the description of the motion of  a  charged  particle on  the  halo is  shown in Section  
\ref{sec: Motion}, whereas  the stability  of  the orbits of  test particles in  the  halo is studied in Section \ref{sect:stability}.
Finally, we complete the paper with a discussion of the results in  Section \ref{sec:conclude}.
  
\section{Exact solutions   for  relativistic magnetized  haloes  surrounding thin disks}\label{sec:formalism}
In this  section  we   consider  the conventional treatment of rotating galaxies modelled as a  stationary  thin disk and,  correspondingly,  we  associate the magnetized halo  
with the region  surrounding the disk. To  do  so,  we formulate  the distributional Einstein-Maxwell field equations assuming axial symmetry \cite{letelier1995space}. We  also  
suppose that the derivatives of the metric and electromagnetic potential across the disk space-like hyper-surface are discontinuous.  To  formulate the corresponding distributional 
form of the  Einstein-Maxwell field equations,   we   introduce the usual  cylindrical coordinates $x^{\alpha} =(t,r,z,\varphi) $ and assume that there exists  an infinitesimally 
thin  disk located  at  the  hypersurface $z=0$. Accordingly, we  identify the  halo surrounding  the  disk with the  positive $(z  \geq 0)$ and  negative $(z  \leq 0)$ regions 
around the equatorial plane $z=0$,  denoted here by the  superscripts ``$\pm$''.  So  that the  metric  and the  electromagnetic  potential can  be written,  respectively, as
$ g_{\alpha\beta} = g^+_{\alpha\beta} \theta (z) + g^-_{\alpha\beta} \{ 1 - \theta (z) \}$
and
$A_{\alpha} = A^+_{\alpha} \theta (z) + A^-_{\alpha} \{ 1 - \theta (z) \}$. Here  $\theta(z)$  denotes the Heaviside  distribution. As a consequence,  the Ricci tensor reads
              \begin{equation}
                 R_{\alpha\beta} = R^+_{\alpha\beta} \theta(z) + R^-_{\alpha\beta} \{ 1 - \theta (z) \} + H_{\alpha\beta} \delta(z), \label{eq:ricdis}
                \end{equation}
where  $\delta (z)$ is the  Dirac distribution and
      \begin{eqnarray}
         H_{\alpha\beta} = \frac{1}{2} \{ \gamma^z_{\alpha} \delta^z_{\beta}  + \gamma^z_{\beta} \delta^z_{\alpha}
        -\gamma^{\mu}_{\mu} \delta^z_{\alpha} \delta^z_{\beta} - g^{zz} \gamma_{\alpha\beta} \},
      \end{eqnarray}
with  $\gamma_{\alpha\beta} =  2{g_{\alpha\beta,z}}$  and all the quantities are evaluated at $z = 0^+$.  In agreement with  (\ref{eq:ricdis}) the energy-momentum tensor and the 
electric current density acquire the  form
          \begin{subequations}
             \begin{eqnarray}
               T_{\alpha\beta} &=& T^+_{\alpha\beta} \theta(z) + T^-_{\alpha\beta} \{ 1 - \theta(z) \} + Q_{\alpha\beta} \delta(z), \label{eq:emtot}\\
                               & \nonumber\\
                  J_{\alpha} &=& J^+_{\alpha} \theta(z) + J^-_{\alpha} \{ 1 - \theta(z) \} + {\cal I}_{\alpha} \delta(z), \label{eq:eccomp}
                    \end{eqnarray}
                     \end{subequations} 
where $T^\pm_{\alpha\beta}$ and $J^\pm_{\alpha}$ are, respectively, the energy-momentum tensors and electric current density of halo. Moreover, $Q_{\alpha\beta}$ and ${\cal 
I}_{\alpha} $ represent the part of the energy-momentum tensor and the electric current density of disk. 
 The energy-momentum tensor  $T^\pm_{\alpha\beta}$ in (\ref{eq:emtot}) is taken to be the sum of two distributional components, the purely electromagnetic (trace-free) part  and a 
``material'' (trace) part, 
         \begin{eqnarray}
                 T^\pm_{\alpha\beta} = E^\pm_{\alpha\beta} + M^\pm_{\alpha\beta},\label{eq:emtotcomp}
              \end{eqnarray}
where $E^\pm_{\alpha\beta}$ is the electromagnetic energy-momentum tensor
            \begin{eqnarray}
               E_{\alpha\beta} = F_{\alpha\nu}F_{\beta}^{\;\,\nu} - \frac{1}{4} g_{\alpha\beta}F_{\mu\nu}F^{\mu\nu}, \label{eq:tab}
               \end{eqnarray}
with $F_{\alpha\beta} =  A_{\beta,\alpha} -  A_{\alpha,\beta}$ and $M^\pm_{\alpha\beta}$  is  an unknown ``material'' energy-momentum tensor  to be obtained. Accordingly, the 
Einstein-Maxwell equations, in geometrized units such that $c = 8\pi G = \mu _{0} = \epsilon _{0} = 1$, are equivalent to the system of equations 
   \begin{subequations} 
        \begin{eqnarray}
        G_{\alpha\beta}^{\pm} =R^\pm_{\alpha\beta}  - \frac{1}{2} g_{\alpha\beta} R^\pm &=& E^\pm_{\alpha\beta}
        +  M^\pm_{\alpha\beta}\label{eq:einspm}\\
        H_{\alpha\beta}  - \frac{1}{2} g_{\alpha\beta} H &=& Q_{\alpha\beta}, \label{eq:einsdis}\\
        F^{\alpha\beta}_{ \pm  \ \ ; \beta}   &=&    J^{\alpha}_\pm ,\label{eq:maxext}\\
       \left [F^{\alpha\beta}\right]n_{_{\beta}} &=&   {\cal I}^{\alpha},\label{eq:emcasj}
     \end{eqnarray}\label{eq:EMequations}\end{subequations}
where $H \equiv g^{\alpha\beta} H_{\alpha\beta}$. The square brackets in expressions such as $[ F^{\alpha\beta}]$ denote the jump of $F^{\alpha\beta}$ across of the surface $z=0$  
and  $n_{_{\beta}}$ denotes a unitary vector in the  direction normal  to it. To  obtain a  solution of  the  distributional Einstein-Maxwell describing  a system  composed  
by a magnetized halo surrounding a  rotating  thin disk   we shall restrict ourselves to  the case where   the  electric potential  $A_t=0$.

To  solve the Einstein-Maxwell equation (\ref{eq:EMequations}) we  assume the conformastationary space-time given  by  the  line element
          \begin{eqnarray}
              ds^2= -e^{2\phi}(dt + \omega d\varphi)^2 + e^{-2\beta\phi} (dr^2 + dz^2 + r^2d\varphi^2) \label{eq:met1},
             \end{eqnarray}
where  $\phi$ depending only  on $r$ and $z$  and $\beta$ is  an arbitrary real constant. We also  assume that the  magnetic potential $A_{\varphi}$ is  time-independent.
Accordingly,  by computing  the Einstein tensor  $G_{\alpha\beta}$ explicitly from the line  element (\ref{eq:met1}) and  the electromagnetic energy-momentum tensor  
(\ref{eq:tab}),  we obtain for  the non-zero components of the energy-momentum tensor  of the  halo (EMTH) $ M^\pm_{\alpha\beta} = E^\pm_{\alpha\beta} - G^\pm_{\alpha\beta}$:
          \begin{subequations}\begin{eqnarray}
              M_{tt}^{\pm}&=&  -e^{2(1 +\beta)\phi}\{ \beta^2 \nabla \phi\cdot\nabla \phi - 2\beta \nabla^2 \phi
                           + \frac{1}{2}r^{-2} e^{2\beta\phi} \nabla A_{\varphi} \cdot \nabla A_{\varphi}
                           - \frac{3}{4}r^{-2}e^{2(1 + \beta)\phi}\nabla \omega\cdot\nabla\omega\}\\
              M_{t\varphi}^{\pm}&=& e^{2(1 +\beta)\phi}\{ \frac{\beta}{2}\nabla\omega\cdot\nabla \phi
                                 +  \frac{3}{4}r^{-2}e^{2(1 +\beta)\phi}\omega \nabla\omega\cdot\nabla\omega
                                 - \beta^2\omega \nabla\phi\cdot\nabla\phi
                                 + 2\beta\omega\nabla^2\phi + \frac{3}{2}\nabla\omega\cdot\nabla\phi\nonumber\\
                                &-& \frac{1}{2}r^{-2}e^{2\beta\phi}\omega\nabla A_{\varphi}\cdot\nabla A_{\varphi}
                                 +\frac{1}{2}\nabla^2\omega- r^{-1}\nabla\omega\cdot\nabla r  \}\\
              M_{rr}^{\pm}&=& (1 - \beta)\nabla^2\phi - (1 - \beta)\phi_{,rr} + (\beta^2 - 2\beta)\phi_{,r}^2
                           + \phi_{,z}^2 -\frac{1}{2}r^{-2}e^{2\beta\phi}(A_{\varphi,r}^2- A_{\varphi,z}^2  )\nonumber\\
                          &+& \frac{1}{4}r^{-2}e^{2(1 + \beta)\phi}(\omega_{,r}^2 - \omega_{,z}^2)\\
              M_{rz}^{\pm}&=& \frac{1}{2}r^{-2}e^{2(1 + \beta)\phi}\omega_{,r}\omega_{,z}
                           - (1 - \beta^2 +2\beta)\phi_{,r}\phi_{,z} - (1 -\beta)\phi_{,rz}
                           - r^{-2}e^{2\beta\phi}A_{\varphi,r}A_{\varphi,z},\\
             M_{zz}^{\pm}&=&-\frac{1}{4}r^{-2}e^{2(1 +\beta)\phi}(\omega_{,r}^2 - \omega_{,z}^2) + \phi_{,r}^2
                          - (1 - \beta)\phi_{,zz} + (1 - \beta)\nabla^2\phi  + (\beta^2 - 2\beta)\phi_{,z}^2\nonumber\\
                        & + & \frac{1}{2}r^{-2}e^{2\beta\phi}(A_{\varphi,r}^2 -A_{\varphi,z}^2),\\
             M_{\varphi\varphi}^{\pm}&=&r^2\nabla\phi\cdot\nabla\phi + (1- \beta)r^2\nabla^2\phi
                                      - (1- \beta)r\nabla\phi\cdot\nabla r
                                      - \frac{1}{2}e^{2\beta\phi}\nabla A_{\varphi}\cdot{\nabla}A_{\varphi}\nonumber\\
                                     & + & 
                   e^{2(1 +\beta)\phi}\Big\{\frac{1}{4}(1  + 3 r^{-2}e^{2\beta\phi}\omega^2)\nabla\omega\cdot\nabla\omega
                   - \beta^2\omega^2\nabla \phi\cdot\nabla \phi + 2 \beta\omega^2\nabla^2\phi
                   + \omega\nabla^2\omega \nonumber\\
                  &-& 2r^{-1}\omega\nabla\omega\cdot\nabla r + (3 + \beta)\omega \nabla\omega\cdot\nabla\phi
                   -\frac{1}{2}r^{-2}e^{2\beta\phi}\omega^2\nabla A_{\varphi}\cdot\nabla A_{\varphi}   \Big\}.
                          \end{eqnarray}\end{subequations}
Moreover,  from (\ref{eq:maxext})  the  non-zero  components  of  the   electric current density  on the  halo have the  form
           \begin{subequations}\begin{eqnarray}
               J^{t}_{\pm}&=& e^{-(1 - 3\beta)\phi} \nabla\cdot \{ \omega r^{-2}e^{ (1 + \beta)\phi} \nabla A_{\varphi}\},\\
                J^{\varphi}_{\pm}&=& e^{-(1 - 3\beta)\phi} \nabla\cdot \{ r^{-2}e^{ (1 + \beta)\phi} \nabla A_{\varphi}  \},
                  \end{eqnarray}\label{eq:haloCURRENT2}\end{subequations}
where  all  the  quantities depending  on $r$ and $z$.       

The  discontinuity in  the  $z$-direction of $Q_{\alpha\beta}$  and  ${\cal I}^{\alpha}$ defines,  respectively,  the surface
energy-momentum tensor (SEMT)  and the surface electric  current density (SECD)  of  the  disk $S_{\alpha\beta}$, more precisely
                    \begin{subequations}\begin{eqnarray}
                        S_{\alpha\beta} &\equiv&   \int Q_{\alpha\beta}  \delta(z)  ds_n  = \sqrt{g_{zz}}  Q_{\alpha\beta}, \label{eq:SEMDn}\\
                              {\cal J}^{\alpha} &\equiv&   \int {\cal I}^{\alpha}  \delta(z)  ds_n  = \sqrt{g_{zz}}  {\cal I}^{\alpha}, \label{eq:SCDn}
                           \end{eqnarray}\end{subequations}
where $ds_n = \sqrt{g_{zz}} \ dz$ is the ``physical measure'' of length in the direction normal to the $z = 0$ surface. Accordingly, for  the metric
(\ref{eq:met1}), the  non-zero components of $S_{\alpha\beta}$ and ${\cal J}^{\alpha}$  are  given by
           \begin{subequations}\begin{eqnarray}
                S_{t t}   &=& 4\beta e^{(2+ \beta )\phi}\phi_{,z} ,\\
                S_{t \varphi}  &=& e^{(2 +\beta)\phi}( 4\beta\omega\phi_{,z} + \omega_{,z}),\\
                S_{r r}  &=& 2(1 - \beta)e^{-\beta \phi}\phi_{,z},\\
                S_{\varphi\varphi}  &=& e^{(2 +\beta)\phi}\{ (4\beta\omega^2
                + 2(1 -\beta)r^2e^{-2(1 + \beta)\phi})\phi_{,z}  +2\omega\omega_{,z}   \},
                \end{eqnarray}\label{eq:SEMTdisk}\end{subequations}
and
             \begin{subequations}\begin{eqnarray}
                 {\cal J}^{t}&=& r^{-2}e^{3\beta \phi}\omega \left[ A_{\varphi,z}\right],\\
                 {\cal J}^{\varphi}&=& -r^{-2}e^{3\beta \phi} \left[ A_{\varphi,z}\right],
                 \end{eqnarray}\label{eq:diskCURRENT2}\end{subequations}
respectively.  Note  that in (\ref{eq:SEMTdisk}) and (\ref{eq:diskCURRENT2})  all the  quantities  are evaluated on the  surface  of  the disk $(z=0)$.

In order  to reduce the  complexity of the last field  equation systems
we  assume   that the  halo's electric current  density  vanishes (i.e.  $ J^{t}_{\pm}=J^{\varphi}_{\pm}=0$ in (\ref{eq:haloCURRENT2})), it  turns  out that the magnetic  
potential  and the metric functions $\phi$ and $\omega $ become  completely determined in terms of an arbitrary harmonic function $U(r,z)$  as  follows (see 
\cite{PhysRevD.87.044010, 2015GReGr.47.54G} for more details),
               \begin{subequations}\begin{eqnarray}
                  A_{\varphi,r} &=&-\frac{1}{k}rU_{,z}, \label{eq:Avarphir}\\
                  A_{\varphi,z} &=&\frac{1}{k}rU_{,r},\label{eq:Avarphiz}\\
                  {(\beta + 1)\phi}&=&-\ln{(1-U) \label{eq:metricpot1}},\\
                  \omega &=& k_{\omega}U\label{eq:metricpot1a},
                  \end{eqnarray}\label{eq:eqitermsofU}\end{subequations}
 with $k$ and $k_{\omega}$  arbitrary constants. Since the non-zero components of the EMTD and EMTH and the electric  current density directly depend on the  metric functions and  
magnetic  potential, we  observe that the  entire solution is determined by a  single harmonic  function.

\section{Exact Relativistic model  for magnetized  disk-haloes}\label{sec:EMT-DH}
So far,  by  using  the  inverse  method and  the  distributional  formulation  of  the  Einstein-Maxwell equations,  we  have  obtained the  separate energy-momentum  tensors of  
the  disk and  halo. In addition,  we  have discussed  out a  method to determinate it  in terms of  an  arbitrary harmonic  function.  Now, the behavior of the  energy-momentum 
tensors obtained  must be investigated   to find what  conditions must be  imposed on the solutions and the parameters that appear in the disk-haloes models  in  such a way that it 
can  describe   reasonably physical  sources.  We  shall now study the  possible  features of  the disk   by  assuming that it is  possible to express its  energy-momentum tensor 
in the 
  canonical  form  
\begin{eqnarray}
               S_{\alpha\beta}&=& (\mu + P)V_{\alpha}V_{\beta} + P g_{\alpha\beta}
                               + {\cal Q}_{\alpha}V_{\beta} + {\cal Q}_{\beta}V_{\alpha}
                               + \Pi_{\alpha\beta}\label{eq:SEMTcanonical},
                 \end{eqnarray}
where  ${\cal Q}_{\alpha}V^{\alpha}={\cal Q}^{\alpha}V_{\alpha}=0$,  $\alpha=(t,r,\varphi)$  and  all the  quantities  are  evaluated  in  $z=0^+$. Similarly,  we assume that its 
possible to   express  the  energy-momentum tensor  of  the halo   in the  canonical  form
        \begin{align}
          M_{\alpha\beta}^{\pm} = (\mu^{\pm} + P^{\pm})V_{\alpha}V_{\beta} + P^{\pm} g_{\alpha\beta}
                               + {\cal Q}_{\alpha}^{\pm}V_{\beta} + {\cal Q}_{\beta}^{\pm}V_{\alpha}
                               + \Pi_{\alpha\beta}^{\pm}\label{eq:EMTHcanonical},
                 \end{align}
where   ${\cal Q}_{\;\;\alpha}^{\pm}V^{\alpha}={\cal Q}^{\pm \alpha}V_{\alpha}=0$, $\alpha=(t,r,z, \varphi)$ and all the  quantities depend on $r$ and $z$. Consequently, we can  
say that  the   disk  and halo are   constituted  by a some mass-energy distributions described by the energy-momentum  tensors (\ref{eq:SEMTcanonical})   and 
(\ref{eq:EMTHcanonical}), respectively.  The $V^{\alpha}$  is the  four velocity of certain observer. Correspondingly, $\mu$, $P$, ${\cal Q}_{\alpha}$ and $\Pi_{\alpha\beta}$ are 
then the  energy density, the isotropic  pressure,  the  heat flux and the anisotropic  tensor on the  surface  of  the  disk. Analogously, 
 $ \mu ^{\pm}$, $  P^{\pm}$, $ {\cal Q}_{\alpha}^{\pm}$ and $\Pi_{\alpha\beta}^{\pm}$ are then the  energy density, the isotropic  pressure,  the  heat flux and the
anisotropic  tensor on the  halo, respectively. Thus, it  is  straightforward to see  that for  the  halo  we  have
      \begin{subequations}\begin{align}
                             \mu ^{\pm}& = M_{\alpha\beta}^{\pm} V^{\alpha}V^{\beta},\label{eq:haloenergy}\\
                                P^{\pm}& =\frac{1}{3}  {\cal H}^{\alpha\beta}M_{\alpha\beta}^{\pm},\label{eq:halopressure} \\
                {\cal Q}_{\alpha}^{\pm}& = -\mu ^{\pm} V_{\alpha} - M_{\alpha\beta}^{\pm} V^{\beta},\label{eq:haloheatflux}\\
                \Pi_{\alpha\beta}^{\pm}& ={\cal H}_{\alpha}^{\;\;\mu}{\cal H}_{\beta}^{\;\;\nu}  ( M_{\mu\nu}^{\pm}  - P^{\pm}{\cal H}_{\mu\nu}),\label{eq:haloanisotropict}
                                          \end{align}\label{eq:hobservables}\end{subequations}
 where the  projection tensor is defined by ${\cal H}_{\mu\nu}\equiv g_{\mu\nu} + V_{\mu}V_{\nu}$ and all  the  quantities depending  on $r$ and $z$.  Whereas ,
for  the   disk  we  have
                                 \begin{subequations}\begin{eqnarray}
                                   \mu &=& S_{\alpha\beta} V^{\alpha}V^{\beta},\label{eq:diskenergy}\\
                                      P&=&\frac{1}{3}  {\cal H}^{\alpha\beta}S_{\alpha\beta},\label{eq:diskpressure} \\
                      {\cal Q}_{\alpha}&=& -\mu V_{\alpha} -S_{\alpha\beta} V^{\beta},\label{eq:diskheatflux}\\
                      \Pi_{\alpha\beta}&=&{\cal H}_{\alpha}^{\;\;\mu}{\cal H}_{\beta}^{\;\;\nu}
                                      ( S_{\mu\nu}  - P{\cal H}_{\mu\nu}),\label{eq:diskanisotropict}
                                          \end{eqnarray}\label{eq:observables}\end{subequations}                                
where all the  quantities  are evaluated  in  $z=0^+$.
It  is easy to  note that  by choosing the angular velocity  to be  zero in  Equation $(\ref{eq:cuadrivelocity})$  we  have then a  fluid comoving in our coordinates  system.  
Hence,  we  
may introduce a  suitable reference  frame in terms  of  the local observers tetrad (\ref{eq:localobservator})  and (\ref{eq:duallocalobservator}) in the  form 
       \begin{eqnarray}
                 \{  V^{\alpha},I^{\alpha},K^{\alpha},Y^{\alpha} \} \equiv     \{h^{\;\;\; \alpha}_{(t)},h^{\;\;\; \alpha}_{(r)},h^{\;\;\; \alpha}_{(z)},h^{\;\;\; 
\alpha}_{(\varphi)}\},
                 \label{eq:loctetrad}
                 \end{eqnarray}
with  the  corresponding dual tetrad 
                \begin{eqnarray}
               \{ V_{\alpha},I_{\alpha},K_{\alpha},Y_{\alpha} \} \equiv     \{-h_{\;\;\; \alpha}^{(t)},h_{\;\;\; \alpha}^{(r)},h_{\;\;\; \alpha}^{(z)},h_{\;\;\; 
\alpha}^{(\varphi)}\}.
                \end{eqnarray}
Since the SECD of  the   disk ${\cal J}^\alpha$ can be also  written  in the  canonical  form
                  ${\cal J}^{\alpha}= { \sigma}V^{\alpha} + {j}Y^{\alpha}  \label{eq:SECDcanonical}$,
$\sigma$ can be  interpreted as  the  surface electric  charge  density and ${j}$  as  the ``current of magnetization'' of the disk. A direct calculation shows that the   
surface electric  charge  density $\sigma =0$,  whereas  the ``current of magnetization'' of the  disk  is  given  by
                   $j =  - r^{-1} e^{2 \beta \phi} \left[  A_{\varphi,z} \right]$,
where,  as  above, $\left[  A_{\varphi,z} \right]$  denotes  the  jump of  the $z-$derivative  of  the  magnetic potential across  of the  disk and,  all quantities  are  
evaluated on the  disk.

By  using the results obtained in the precedent section, we can write  the  surface energy density of  the  disk and  the  energy  density of the  halo  can written as
\begin{align}
\mu(r) =  \frac{ 4\beta U_{,z} }
                              { (\beta + 1) (1-U)^{\frac{2\beta + 1}{\beta +1}} }\label{eq:diskenergy+U}
\end{align}
and
\begin{align}
                        \mu ^{\pm}(r,z)  = \frac{(U_{,r}^2+ U_{,z}^2)e^{2(1+2\beta)\phi}}{(1 + \beta)^2r^2}
                           \left\{ (2\beta+ \beta^2)r^2 - \frac{(1+ \beta)^2}{2k^2}r^2 e^{-2\phi} + \frac{3k_{\omega}^2(1+ \beta)^2}{4} \right\},
                            \end{align}
respectively. Moreover,  we  have a  barotropic equation of  state on the  surface  of  the  disk, which  can be  given  by   $P(r)=\eta\mu\ $,  with ${ \eta} = {(1 -\beta)}/{3 
\beta}$, in such a way that  we  can  use  the energy  conditions   and  the  causality requirement for  the speed  of  sound on  the  disk  to  obtain  the physical range of  
possible values  of the  parameter $\beta$. Analogously,  the  pressure of the  halo  we  have  $P^{\pm}(r,z) = \Theta \mu ^{\pm}(r,z) $, where
\begin{align*}
 \Theta:= \frac{  (4- 2\beta -  \beta^2)r^2  - \frac{(1+ \beta)^2}{2k^2}r^2 e^{-2\phi}
                 + \frac{k_{\omega}^2(1+ \beta)^2}{4} \left( 1 + 3k_{\omega}^2r^{-2}U^2 e^{2\beta\phi}(1 - e^{2\phi})  \right)  }
                 {3 \left( (2\beta+ \beta^2)r^2 
- \frac{(1+ \beta)^2}{2k^2}r^2 e^{-2\phi} + \frac{3k_{\omega}^2(1+ \beta)^2}{4} \right)},
\end{align*}
in such a way that the pressure of  the  halo not only depends on the energy density but also on the gravitational and magnetic fields through the function $\Theta$. The  Heat  
function  of  the  disk  is  given by
\begin{align}
  {\cal Q}_{\alpha}(r)=-\frac{k_{\omega} U_{,z}}{1-U} \delta_{\alpha}^{\;\varphi},\label{eq:diskheatflux+U}
\end{align}
Similarly,  by  inserting   (\ref{eq:EMTHcanonical}) into   (\ref{eq:hobservables})    we  obtain  for  the  heat flux of  the  halo
                \begin{align}
                  {\cal Q}_{\alpha}^{\pm}&= \frac{k_{\omega}e^{(1 + 2\beta)\phi}}{2(1 + \beta)r}
                                                                      \left\{2(1 + \beta)U_{,r} - (3 + \beta)r(U_{,r}^2  +  U_{,z}^2 )e^{(1+\beta)\phi} \right\}  
                                                                      \delta^{\varphi}_{\alpha}.\label{eq:diskheatfluxExplicit}
                         \end{align}
The  non-zero  components  of  the  anisotropic tensor  of  the  disk read $\Pi_{\varphi\varphi}(r)= r^2 \Pi_{rr} \label{eq:diskanisotropictvv+U}(r)$ where 
\begin{align}
 \Pi_{rr}(r)=\frac{2(1 -\beta)U_{,z}}{3(1+ \beta)(1-U)^{\frac{1}{1+\beta}}}\label{eq:diskanisotropictrr+U}.
\end{align}
Moreover, it is easy  to  see  that   the  anisotropic tensor of  the  halo reads
            \begin{align}
             \Pi_{\alpha\beta}^{\pm}  =     P_{r}^{\pm} I_{\alpha} I_{\beta} 
                                                                +  P_{z}^{\pm} K_{\alpha} K_{\beta}
                                                                +  P_{\varphi}^{\pm} Y_{\alpha} Y_{\beta} 
                                                                +  2P_{T}^{\pm}  I_{(\alpha} K_{\beta)}
                    \end{align}
where 
                \begin{align}
                       P_{r}^{\pm}&= e^{2\beta \phi} \Pi_{rr}^{\pm},\\
                       P_{z}^{\pm}&= e^{2\beta \phi} \Pi_{zz}^{\pm},\\
         P_{\varphi}^{\pm} &=  \frac{e^{2\beta \phi}}{r^2} \Pi_{\varphi\varphi}^{\pm},\\
                      P_{T}^{\pm}& = e^{2\beta \phi}    \Pi_{rz}^{\pm}.
                     \end{align}
and
        \begin{subequations}\begin{align}
         \Pi_{rr}^{\pm} &= \frac{e^{2(1+\beta)\phi}}{3(1 + \beta)^2r^2}
                                     \bigg \{
                                         \bigg(   \frac{k_{\omega}^2 (1 + \beta)^2}{2} + \frac{2(1+ \beta)^2 }{k^2}r^2e^{-2\phi} - 4 (1 + \beta - \beta^2)r^2
                                                -  \frac{3k_{\omega}^4 (1+ \beta)^2}{4}r^{-2}U^2 e^{2\beta\phi}(1 - e^{2\phi})     \bigg) U_r^2     \nonumber \\
                                    &  + \bigg(     -  k_{\omega}^2 (1 + \beta)^2          -  \frac{  (1+ \beta)^2 }{k^2}r^2e^{-2\phi}  +2 (1 + \beta - \beta^2)r^2
                                          -  \frac{3k_{\omega}^4 (1+ \beta)^2}{4}r^{-2}U^2 e^{2\beta\phi}(1 - e^{2\phi})  \bigg) U_z^2   \nonumber\\
                                    &   -   3(1 -\beta^2)r^2e^{-(1+ \beta)\phi}U_{,rr}
                                          \bigg \}, \\
        \Pi_{zz}^{\pm}&  =\frac{e^{2(1+\beta)\phi}}{3(1 + \beta)^2r^2}
                                      \bigg \{
                                           \bigg(          - {k_{\omega}^2 (1 + \beta)^2}     -   \frac{  (1+ \beta)^2 }{k^2}r^2e^{-2\phi} + 2  (1 + \beta - \beta^2)r^2
                                            -  \frac{3k_{\omega}^4 (1+ \beta)^2}{4}r^{-2}U^2 e^{2\beta\phi}(1 - e^{2\phi})  \bigg)U_r^2
                                                \nonumber\\
                                         &  + \bigg(      \frac{k_{\omega}^2 (1 + \beta)^2 }{2} + \frac{2(1+ \beta)^2 }{k^2}r^2e^{-2\phi}  - 4 (1 + \beta - \beta^2)r^2
                                          -  \frac{3k_{\omega}^4 (1+ \beta)^2}{4}r^{-2}U^2 e^{2\beta\phi}(1 - e^{2\phi}) \bigg)U_z^2
                                         \nonumber\\
                                       &  - 3(1 -\beta^2)r^2e^{-(1+ \beta)\phi}U_{,zz}
                                                 \bigg \}, \\
  \Pi_{\varphi\varphi}^{\pm}&= \frac{(U_{,r}^2+ U_{,z}^2)e^{2(1+\beta)\phi}}{3(1 + \beta)^2}
                                         \bigg\{
                                             2(1 +  \beta -  \beta^2)r^2  - \frac{(1+ \beta)^2}{k^2}r^2 e^{-2\phi}
                                            + \frac{k_{\omega}^2(1+ \beta)^2}{2} \left( 1 + 3k_{\omega}^2r^{-2}U^2 e^{2\beta\phi}(1 - e^{2\phi})  \right)
                                                \bigg\}
                                                \nonumber\\
                                      & - \frac{(1-\beta)}{1+ \beta}re^{(1+ \beta)\phi}U_{,r},\\
           \Pi_{rz}^{\pm} & =  \frac{e^{2(1 + \beta \phi)}}{(1 + \beta)^2}
                                \bigg\{
                          \bigg( -2(1 + \beta - \beta^2) + \frac{(1 + \beta)^2}{k^2}e^{-2\phi} +\frac{k_{\omega}^2(1 + \beta)^2}{2r^2}  \bigg) U_{,r}U_{,z}
                        - (1 -\beta^2) e^{-(1 + \beta)\phi} U_{,rz}               \bigg \}.
             \end{align}\label{eq:anisotropictensor}\end{subequations}
Notice that  ${\cal P}^{\pm} \equiv P_r^{\pm} + P_z^{\pm} + P_{\varphi}^{\pm}=0 $ and, consequently, the trace    $\Pi_{\quad\alpha}^{\pm \alpha} =0$. We have  obtained
 expressions for  the energy,  pressure and  the other quantities characterizing the dynamic  of the  halo.   All  the dynamic quantities have been expressed  in 
terms of  an arbitrary $U(r,z)$ harmonic  function. Finally,  as  we  know,  the  electric current  density of the  halo is  zero whereas it is easy to note that the  
magnetization current density on surface of the  disk  is
\begin{align}
j(r)=-\frac{\left[ U_{,r}\right]  }{k (1-U)^{\frac{3\beta}{1+\beta}}} \label{eq:diskcurrent+U}.
\end{align}

It is  important  to remark   that  $k_{\omega}$ is a  defining  constant in  (\ref{eq:diskheatflux+U}) and (\ref{eq:diskheatfluxExplicit}).  Indeed,  when $k_{\omega}=0$  the  
heat flux functions $ {\cal Q}_{\alpha}$  and $ {\cal Q}_{\alpha}^\pm$ vanish, a feature of  the  static  systems.  Due to we used the inverse method, no   ``a priori''  
restriction are imposed on the physical properties of  the material constituting the  disk  and  halo.  The  non-zero components of  the energy-momentum tensors of the   disk and 
halo  result of ``the nature'' of  the  chosen metric and  the corresponding solutions. So, in our case, the non-zero component $S_{rr}$ and  $S_{t\varphi}$  
are conditioned by the parameter $\beta$ and  the  metric function $\omega$ in such a way  that   when $\beta =1$  the component $S_{rr}$ vanishes, whereas $S_{t\varphi}=0$   
when $\omega$ vanishes.  The decomposition of the  energy-momentum  tensor  of the  disk-halo system into (\ref{eq:SEMTcanonical}) and (\ref{eq:EMTHcanonical})   were  
chosen  with the  aim  to describe  the SEMT and  EMTH by   the  more general  fluid model.  Hence, the  heat flux appear here in a  ``natural''  way as  a function determined by 
the metric function $\omega$ and, consequently, by  the  ``rotation''. Unfortunately,  as  we  can  see  from (\ref{eq:diskheatflux})  and (\ref{eq:haloheatflux}),  this function  
is  oriented along the closed  circular orbits and thus   its physical  interpretation is  unclear. It  is an issue that remains unanswered in this manuscript, but  should be 
addressed in the  future.

\section{Rotating Kuzmin-like disk with magnetized  haloes}\label{sec:examples}
 \begin{figure}[h]
$$\begin{array}{cc}
\includegraphics[width=0.5\textwidth]{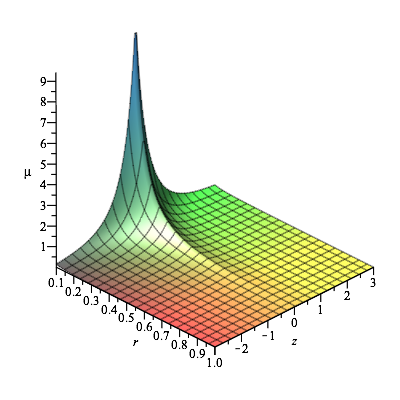}&
\includegraphics[width=0.5\textwidth]{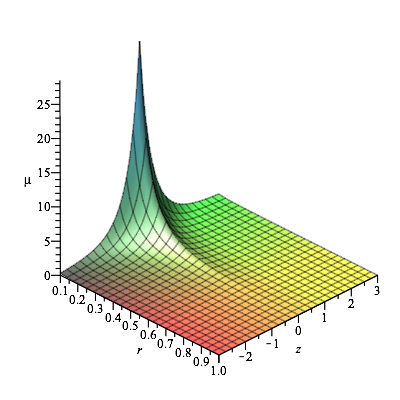}\\
 (a) & (b) \\
 \end{array}$$
\caption{\label{fig:figure1a}Surface  plots    of  the  energy  density
 (a)  ${\mu}^{\pm}_{_0}$   and (b)  ${\mu}^{\pm}_{_1}$  on  the   halo   as a functions depending on  ${ r}$  and   ${z}$ with  parameters 
   $a={b}_0={b}_1= k=k_{\omega}=1$ and $\beta =0.75$.}
\end{figure}

 \begin{figure}[h]
$$\begin{array}{cc}
\includegraphics[width=0.5\textwidth]{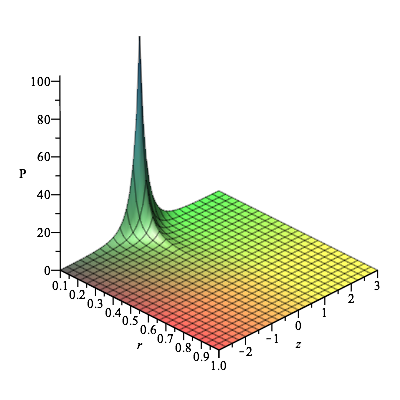}&
\includegraphics[width=0.5\textwidth]{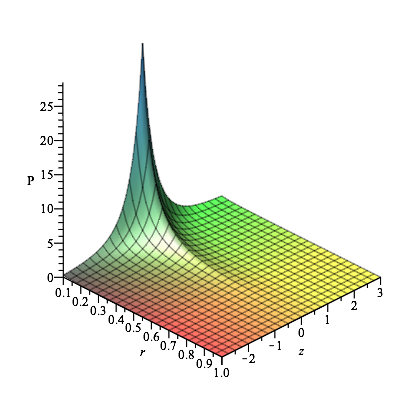}\\
 (a) & (b) \\
 \end{array}$$
\caption{\label{fig:figure2a}Surface  plots    of  the  pressure
 (a)  ${P}^{\pm}_{_0}$   and (b)  ${P}^{\pm}_{_1}$  on  the  halo   as a functions depending on  ${ r}$  and   ${z}$ with  parameters 
   $a={b}_0={b}_1= k=k_{\omega}=1$ and $\beta =0.75$.}
\end{figure}
As  an example  of  application  of the  formalism described  in the precedent sections,  we  now consider the   magnetized  haloes surrounding  the  rotating  disks  generated  
by a generalization of  the Kuzmin-disk potential in the  form \cite{binney1998galactic, stephani2003exact}
              \begin{eqnarray}
                    U= - \sum_{n=0}^{N}{\frac{b_n P_n(z/R)}{R^{n + 1}}} ,
                 \qquad P_n(z/R)= (-1)^n\frac{R^{n+1}}{n!}\frac{\partial^n}{\partial z^n}\left(\frac{1}{R}\right)
                    \label{eq:GeneralizedKuzmin},
                  \end{eqnarray}
where $P_n=P_n(z/R)$  is the  Legendre polynomials  in cylindrical  coordinates that was derived in the present form by a direct comparison of    the  Legendre 
polynomial expansion of   the  generating function with  a  Taylor of $1/R$ \cite{arfken1966mathematical}, being $R^2\equiv r^2 + z^2$ and  $b_n$  arbitrary constant  
coefficients. The corresponding magnetic  potential,  obtained  from (\ref{eq:Avarphir}) and (\ref{eq:Avarphiz}), is
            \begin{eqnarray}
                  A_{\varphi} =- \frac{1}{k} \sum_{n=0}^{N} b_n \frac{(-1)^n}{n!}\label{eq:magpot}
                   \frac{\partial ^n}{\partial z^n}\left(\frac{z}{R}\right)
                  \end{eqnarray}
where,    we  have imposed  $A_{\varphi}(0,z)=0$ in order  to  preserve the regularity of the axis of  symmetry. Next,  to introduce  the corresponding  discontinuity  in the  
first-order derivatives of  the metric potential and  the magnetic  potential required  to  define the disk we  perform  the  transformation $z \rightarrow |z| +  a$. 
It is worth noting that, for  the  two  first members of  the  family of  solutions ($N=0$ and $N=1$) we  have
           \begin{subequations}\begin{eqnarray}
          U_0 &=& - \frac{\tilde{b}_{_0}}{\sqrt{   \tilde{r}^2 +  (|\tilde{z}| + 1)^2}},\label{eq:GeneralizedKuzmin0}\\
          \tilde{A}_{\varphi 0} &=&-\frac{\tilde{b}_{_0}(|\tilde{z}|+1)}{k \sqrt{ \tilde{r}^2 +(|\tilde{z}| + 1)^2}},\label{eq:magpot0}.
             \end{eqnarray}\end{subequations}
 and       
          \begin{subequations}\begin{eqnarray}
          \tilde{A}_{\varphi 1} &=& -\frac{\tilde{b}_{_0} (|\tilde{z}| + 1)}{k \sqrt{ \tilde{r}^2 + (|\tilde{z}| + 1)^2}}
          \left\{ 1 - \frac{\tilde{b}_{_1}\tilde{r}^2 }
          {\tilde{b}_{_0} (|\tilde{z}| + 1) \left((|\tilde{z}| + 1)^2 + \tilde{r}^2 \right)}    \right\}\label{eq:magpot1},\\
          U_1 &=& - \frac{\tilde{b}_{_0}}{\sqrt{   \tilde{r}^2 +  (|\tilde{z}| + 1)^2}}
              \left\{1 +\frac{\tilde{b}_{_1} (|\tilde{z}| +1)}{\tilde{b}_{_0}\left((|\tilde{z}| + 1)^2 + \tilde{r}^2 \right)} \right\}\label{eq:GeneralizedKuzmin1},            
            \end{eqnarray}\end{subequations}
respectively, where $\tilde{b}_{_0} = {b}_{_0}/a$ and $\tilde{b}_{_1} = {b}_{_1}/a^2$ whereas $ \tilde{r}=r/a$ and  $\tilde{z}=z/a$, moreover  $\tilde{A}_{\varphi}=   
A_{\varphi}/a$. For the  two  first  members of  the  family  of  solutions   the  surface energy  density of  the  disks  can  be  written  as 
           \begin{eqnarray}
               \tilde{\mu}_0 = \frac{4  \beta  {\tilde b}_{_0}}
                      {(1 + \beta)(\tilde{r}^2 + 1)^{(\beta + 2)/(2 \beta + 2)}   \left( \tilde{b}_{_0}
                      + \sqrt{\tilde{r}^2 + 1}  \right)^{(2\beta + 1)/(\beta + 1)}},
                      \end{eqnarray}
                      
and                      
                      \begin{eqnarray}
      \tilde{\mu} _1=\frac{4 \beta \{({\tilde b}_{_0}- {\tilde b}_{_1})\tilde{r}^2+{\tilde b}_{_0} + 2 {\tilde b}_{_1}\}}
                      {(1 + \beta)(\tilde{r}^2 + 1)^{(2  -\beta)/(2 \beta + 2)}
   \left\{ (\tilde{r}^2 + 1)^{3/2}+{\tilde b}_{_0} (\tilde{r}^2 + 1)+\tilde{b}_{_1}\right\}^{(2\beta +1)/(\beta +1)} } .
                 \end{eqnarray}
respectively. Similarly, for the  two first members of  the  family we have  the  heat  flux of  the  disks
            \begin{subequations}\begin{eqnarray}
                 {Q}_{\alpha 0} &=& \frac{ \tilde{b}_{_0} k_{\omega} \delta_{\alpha}^{\varphi} } { \sqrt{\tilde{r}^2 + 1} +  \tilde{b}_{_0}},\\
                 {Q}_{\alpha 1} &=& \frac{ k_{\omega}\delta_{\alpha}^{\varphi} \left(\tilde{b}_{_0}(\tilde{r}^2 + 1) + \tilde{b}_{_1} \right) }
             {(\tilde{r}^2 + 1)^{3/2} + \tilde{b}_{_0} (\tilde{r}^2 + 1) + \tilde{b}_{_1} },
                \end{eqnarray}\end{subequations}
and  the  corresponding anisotropic tensor
                   \begin{subequations}\begin{eqnarray}
                   \tilde{\Pi}_{rr 0}  &=& \frac{2(1 - \beta)\tilde{b}_{_0}}
                  {3(1 +\beta) \left( \sqrt{\tilde{r}^2 + 1}  +  \tilde{b}_{_0} \right)^{1/(1+ \beta)}\left(\tilde{r}^2 + 1\right)^{(3\beta + 2)/(2 + 2\beta)} },\\
                   \tilde{\Pi}_{rr 1}  &=& \frac{2(1 - \beta)\left((\tilde{b}_{_0} - \tilde{b}_{_1})\tilde{r}^2 + \tilde{b}_{_0} + 2\tilde{b}_{_1}  \right)  }
                  {3(1 +\beta) \left\{ (\tilde{r}^2 + 1)^{3/2} +  \tilde{b}_{_0} (\tilde{r}^2 + 1)  +    \tilde{b}_{_1}\right\}^{1/(1+ \beta)}
                  \left(\tilde{r}^2 + 1\right)^{(5\beta + 2)/(2 + 2\beta)} }.
                   \end{eqnarray}\end{subequations}
 As  we know, the another  quantities  are  $P=(1-\beta)\mu/(3\beta)$  
and  $ \Pi_{\varphi\varphi} =r^2 \Pi_{rr} $.
In the  last expressions we  have  used  the dimensionless expressions $\tilde{\mu}=a \mu $, $\tilde{\Pi}_{\varphi\varphi}=a\Pi_{rr}$
Finally, for the  two first members of  the  family we have the   current of magnetization as
       \begin{subequations}\begin{eqnarray}
           \tilde{j_0} &=& -\frac{ 2\tilde{b}_{_0} \tilde{r}}
                            { k (\tilde{r}^2 + 1)^{(3+\beta)/(2 + 2\beta)}\left( \tilde{b}_{_0}
                                           + \sqrt{\tilde{r}^2 + 1}   \right)^{2\beta/(\beta + 1)}} ,\\
                \tilde{j_1} &=& -\frac{ 2\tilde{r} \left( \tilde{b}_{_0} (\tilde{r}^2 + 1) + 3\tilde{b}_{_1} \right) }
                            {k (\tilde{r}^2 + 1)^{(5-\beta)/(2 + 2\beta)}\left\{ (\tilde{r}^2 + 1)^{3/2} +  \tilde{b}_{_0} (\tilde{r}^2 + 1)
                            +    \tilde{b}_{_1}\right\}^{2\beta /(1+ \beta)}   }.
        \end{eqnarray}\end{subequations}
where
$\tilde{j}=aj$ and  we first have assumed that the  $z$-derivative  of  the  magnetic
potential  present  a finite  discontinuity through  the disk.  In fact, as  we  have  said  above,  the derivatives  of
 $U$ and  $A_{\varphi }$  are  continuous functions across  of  the  surface  of  the disk.    We  artificially
introduce the  discontinuity through  the  transformation     $z \rightarrow |z| +  a$ .

It  is  worth noticing that  the  mass surface  density  as well  as the  isotropic pressure of  the  disk decay very rapidly (as  $1/r^3$  and $1/r^5$ for $N=0$ and $N=1$, 
respectively)  indicating that the  above  solution  can  be  associated with a disk   with  a finite energy-momentum distribution. In  every  case,  the  characteristic  size  
can  be adjusted  through the  parameters  $b_0$ and $b_1$  of the  solutions. Moreover,  a simple  calculation  of the curvature invariants reveals that  the  solution  is  
asymptotically flat and  singularity-free.

To  illustrate the results corresponding  to the principal quantities describing the   halo in Fig. 
\ref{fig:figure1a}, we show the behavior of   energy  densities  ${{\mu}^{\pm}}$ on  the halo  as a function of $r$ and $z$. In each case, we plot ${\mu}^{\pm}_0(r,z)$  [Fig. 
\ref{fig:figure1a}(a)]  and ${\mu}^{\pm}_1(r,z)$  [Fig. \ref{fig:figure1a}(b)] for the indicate values of the parameters. It can be seen that the energy density is everywhere 
positive and  vanishes sufficiently fast as $r$ increases.

In Fig. \ref{fig:figure2a}, we show the behavior of   pressure  ${P^{\pm}}$ on  the halo  as a function of $r$ and $z$.  In each case, we plot ${P}^{\pm}_0(r,z)$ [Fig. 
\ref{fig:figure2a}(a)]  and ${P}^{\pm}_1(r,z)$  [Fig. \ref{fig:figure2a}(b)] for the indicate values of the parameters. We can see that pressure is always positive and behaves as 
the energy density of the  halo. Thus, we can see that the  behavior of  these quantities are in agreement with the results published in \cite{Chakraborty:2014paa}. 
Moreover, we also computed these functions for other values of the parameters within the allowed range and in all cases we have found a similar behavior.

\subsection{The constants of  motion}
To proceed further,  we  evaluate the  constants  of  motion.  Therefore,  from  (\ref{eq:metricpot1})  we have
         \begin{eqnarray}
                  \phi= \frac{1}{1+ \beta} \ln{\left( \frac{1}{1 - U} \right)}.\label{eq:phi}
          \end{eqnarray}
Then,  for  the solution (\ref{eq:GeneralizedKuzmin0}) we may  write
         \begin{eqnarray}
                  \phi_0= \frac{1}{1+ \beta}
                             \ln{ \left(  \frac{  \sqrt{ \tilde{R}^2 +  2 |\tilde{z}|  + 1}}{  \sqrt{ \tilde{R}^2 +  2 |\tilde{z}|  + 1}   +  {\tilde{b}}_{_{0}}}   \right) }, 
\label{eq:phi0}                         
          \end{eqnarray}
where $\tilde{R}^2 \equiv \tilde{r}^2 + \tilde{z}^2$.         
This follows that  the metric  potentials $ g_{tt}$ and $g_{t\varphi} $ for  $R  \rightarrow \infty$ in  the disk ($z=0$ ) become
      \begin{subequations}    
         \begin{eqnarray}
               g_{tt}             &\simeq&  -1 +  \frac{2 \tilde{b}_{_{0}}}{(1 + \beta) \tilde{R}}  -   \frac{  \tilde{b}_{_{0}}^2 (3 + \beta) }{(1 + \beta)^2 \tilde{R}^2} 
                                                     + O\left( \frac{1}{ \tilde{R}^3 } \right),\\
               g_{t\varphi}   &\simeq&    \frac{k_{\omega} \tilde{b}_{_{0}}}{ \tilde{R} }   - \frac{ 2k_\omega  \tilde{b}_{_{0}}^2}{(1 + \beta) \tilde{R}^2}    
                                                     + O\left( \frac{1}{ \tilde{R}^3 } \right).
             \end{eqnarray}
             \end{subequations}
This  implies,  as  is  well known (See \cite{katz1999disc}),  that the  total mass-energy of space-time associated  with  the disk is 
             \begin{eqnarray}
             M_0= \frac{ {b}_{_{0}} }{ (1 + \beta)} .
               \end{eqnarray}
On the  other  hand,  in $(x,y,z)$ coordinates we  find that 
               \begin{subequations}    
                   \begin{eqnarray}
                      g_{01}             &\simeq&   - \frac{k_\omega \tilde{b}_{_{0}} y}{ \tilde{R}^3} + O\left( \frac{1}{ \tilde{R}^4 } \right), \\
                      g_{02}              &\simeq&   \frac{k_\omega \tilde{b}_{_{0}} x}{ \tilde{R}^3} + O\left( \frac{1}{ \tilde{R}^4 } \right),    \\
                      g_{03}             &\simeq& O\left( \frac{1}{ \tilde{R}^4 } \right).
                 \end{eqnarray}
             \end{subequations}
As an application, we use the same procedure as in \cite{carmeli1982classical}  and   see that the  angular momentum $L_{M0}$ is  in the  $z$-direction  and is  
given by
                 \begin{eqnarray}
                           L_{M0} = \frac{1}{2} k_{\omega} {b}_{_{0}} .
                           \end{eqnarray}
According  to  (\ref{eq:magpot0})  the  magnetic  field is 
                   \begin{eqnarray}
                   \mathbf{B}_0= - \frac{ \tilde{b}_{_{0}} \tilde{r} }{ k (  \tilde{R}^2 +  2 |\tilde{z}|  + 1  )^{3/2}}
                                            \Big( \tilde{r} \mathbf{e}_r    +     \Big( \tilde{z} + \frac{|\tilde{z}|}{\tilde{z}} \Big) \mathbf{e}_z    \Big),
                    \end{eqnarray}
where $ \mathbf{e}_{\alpha}$  are  unit basis vectors in cylindrical  coordinates. Accordingly,  by  expressing the components of the magnetic  field in  
Cartesian coordinates and   taking the   limit  as $R \rightarrow  \infty$  of  $\mathbf{B}_0 (x,y,z)$    and  by using the  formula    (44.4)  of  Landau and Lifshitz 
\cite{landau1975classical} we  may conclude that the magnetic  momentum may  be  written as 
                 \begin{eqnarray}
                           L_{B0} = \frac{ {b}_{_{0}} }{k}. 
                           \end{eqnarray}
We thus  see that constants $k$  and $k_\omega$  defines the  gyromagnetic ratio $ L_{M0}/ L_{B0} = (k k_{\omega})/2$.

\section{Motion of  a charged test particle in the  halo}\label{sec: Motion}
 \begin{figure}[h]
$$\begin{array}{cc}
\includegraphics[width=0.5\textwidth]{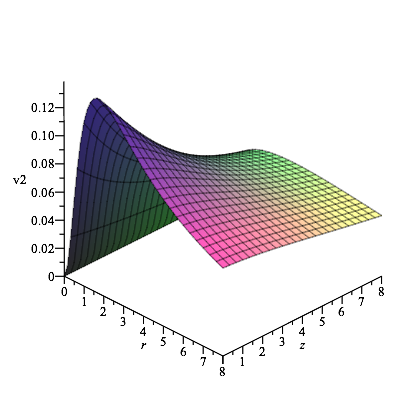}&
\includegraphics[width=0.45\textwidth]{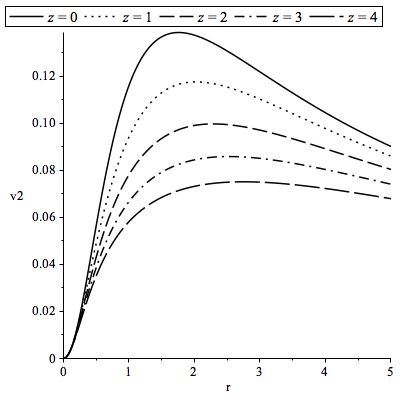}\\
 (a) & (b) \\
 \end{array}$$
\caption{\label{fig:figure51} Surface  plot    of  the  velocity (a)  ${v}^{ 2}_{_0}$   and $z$-slices   of the surface 
plot of the  velocity  (b)    on  the  halo   as a functions depending on  ${ r}$  and   ${z}$ 
with  parameters 
   $a={b}_0={b}_1= k=k_{\omega}=1$ and $\beta =0.75$.}
\end{figure}
 \begin{figure}[h]
$$\begin{array}{cc}
\includegraphics[width=0.5\textwidth]{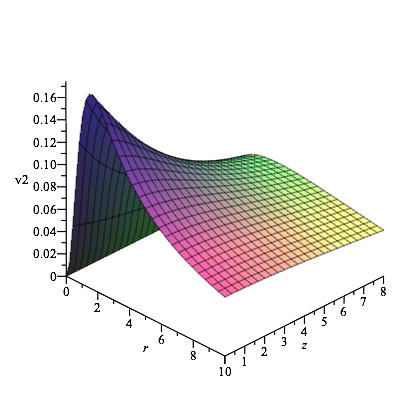}&
\includegraphics[width=0.45\textwidth]{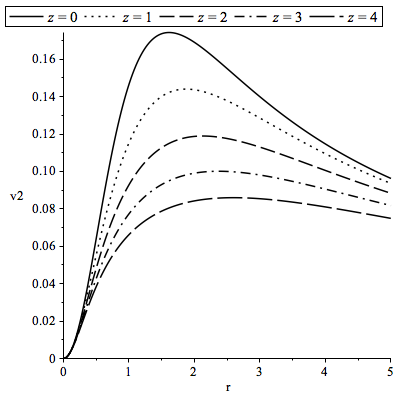}\\
 (a) & (b) \\
 \end{array}$$
\caption{\label{fig:figure71}Surface  plot    of  the  velocity
(a)  ${v}^{ 2}_{_1}$   and $z$-slices   of the surface plot of the  velocity (b) on the  halo   as a functions depending on  ${ r}$  and   ${z}$ with  parameters 
   $a={b}_0={b}_1= k=k_{\omega}=1$ and $\beta =0.75$.}
\end{figure}
The  motion of  a  test  particle of   charge $e$  and  mass $m$ moving in the halo 
 is described by the following Lagrangian density:
\begin{align}
{\cal L}= \frac{1}{2}g_{\alpha\beta} v^{\alpha} v^{\beta} 
        + \frac{e}{m}A_{\alpha}x^{\alpha},\label{eq:lag}
\end{align}
where  $g_{\alpha\beta}$  and $A_{\alpha}$  are, respectively, the components of the  metric and electromagnetic potential, given here  by 
(\ref{eq:eqitermsofU}). 

The equations of motion of the test particle can be derived from  (\ref{eq:lag})
by using the Euler-Lagrange equation. Then,
                     \begin{align}
                         \frac{dv^{\alpha}}{ds} + \Gamma^{\alpha}_{\beta\gamma}v^{\beta}v^{\gamma}
                                                                        = \frac{e}{m}g^{\alpha\mu}F_{\mu\lambda}v^{\lambda}.
                         \end{align}
The   velocity of  the  particle as measured  by  the local observers 
 is  given  by  $v^{\alpha}=v^t(t^{\alpha }+ \Omega \varphi^{\alpha})$,  where
                         \begin{align}
                            v^t &=\frac{(1 -U)^{1/(1+ \beta)}}{(1 + k_{\omega}U\Omega)\sqrt{1 - v^2}}.
                         \end{align}
Here, the 3-velocity $v$ and  the  angular  velocity $\Omega$  of  the  particle as measured  by  the local observers  are given by
         \begin{align}
              v&=\frac{r \Omega (1 -U)}{1 + k_{\omega}U\Omega}
                      \end{align}
and              
             \begin{align}
              \Omega&=\frac{k_{\omega}(U_{,r}^2 + U_{,z}^2) \left((1 + \beta)   + \frac{2U}{1 -U}\right) \pm \sqrt{(U_{,r}^2 + U_{,z}^2) D}}
                                        {2(1 +  \beta)r (1 - U)^2U_{,r}  - 2 (U_{,r}^2 + U_{,z}^2)  A},\\
                       D&=  4(1 + \beta)r(1 - U)U_{,r} + (U_{,r}^2 + U_{,z}^2) \left( k_{\omega}^2(1 + \beta)^2   - 4\beta r^2     \right) ,\nonumber\\
                       A&= \beta r^2(1 - U) + k_{\omega}^2U \left( 1 + \beta + \frac{U}{1 -U}   \right),\nonumber
              \end{align}
respectively.   All the  quantities  depend on $r$ and $z$. In Fig. \ref{fig:figure51}(a) and Fig. \ref{fig:figure51}(a) we show the behavior of  the velocity ${v^2}_0$  and  
${v^2}_1$ of  a charged  particle    following an ``magnetogeodesic'' motion on  the  halo  for  the  values of  indicated  parameters, respectively.  Additionally,  in  Fig. 
\ref{fig:figure71}(b) and Fig. \ref{fig:figure71}(b),   we  plot the $z$-slices   of the surface plot of the  velocity and ${v^2}_0$ and ${v^2}_1$  for the indicated values of the 
parameters, respectively. These  curves are obtained  via vertical slices  of the surface $v^2=v^2(r,z)$ (a vertical slice is a curve formed by the intersection of the surface 
$v^2=v^2(r,z)$ with the vertical planes).  For  each  curve,  we  can  see that the velocity is  always less than 1, its maximum  occurs around  $r=0$, and it vanishes 
sufficiently 
fast as $r$ increases. It can also be observed that the maximum of the velocity decreases as the values of  $z$ increases. We also computed these functions for other values of the 
parameters within the allowed range and in all cases we found a similar behaviour. Naturally, the description of the  motion of charged  particles on disk  here deduced  is  in 
agreement with the  results of analysis of the magnetogeodesic motion of the particle in the  magnetized  disks discussed  in \cite{garcia2014exact}.

\section{Stability  of  orbits of  particles in  the  halo}\label{sect:stability}
Since  the  Lagrangian density (\ref{eq:lag})  does not  depend explicitly on  variables $t$ and 
$\varphi$,  the  following two  conserved quantities exist
\begin{subequations}\begin{align}
p_t= \frac{\partial{\cal L}}{\partial \dot{t}}
= -\frac{E}{m},\\
p_{\varphi}= \frac{\partial{\cal L}}{\partial \dot{\varphi}}= \frac{L}{m},
\end{align}\end{subequations}
where $L$  and $E$  are,  respectively,  the  angular momentum  and  energy of  the particle as measured by 
an observer at rest at  infinity. In the halo near the surface  of  the  disk $(z \rightarrow 0^+)$ 
the motion equations can be reduced to the form $\dot{r}^2 + V(r)= E^2/m^2$, which describes the  motion inside  an
effective  potential $V$ given by
\begin{align}
V(r)=\frac{1}{m^2r^2 (1 - U)^{\frac{4\beta}{1 +\beta}}}
\Bigg[ 
{E^2r^2}\Big(  (1 - U)^{\frac{4\beta}{1 +\beta}} + \frac{m^2}{E^2} (1 - U)^{\frac{2\beta}{1 +\beta}}
- (1-U)^2 \Big) + \Big( {E k_{\omega} U} + {L} \Big)^2
\Bigg],
\end{align}
where,  as  we  know, $U$ is an arbitrary harmonic solution. 
Circular  orbits  are  defined  by $r= r_c =\text{constant}$, so  that $\frac{dr_0}{ds}=0$ and, additionally,
$\frac{dV}{ds}|_{r=r_0}=0$. From these  two  conditions  it's possible  to  evaluate the  conserved quantities $E$
and  $L$. The  orbits will be  stable if  $\frac{d^2V}{ds^2}|_{r=r_0} < 0$ and unstable if $\frac{d^2V}{ds^2}|_{r=r_0} > 0$.

\begin{figure}
$$\begin{array}{cc}
\includegraphics[width=0.4\textwidth]{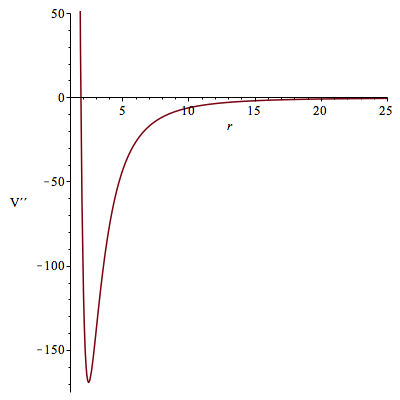}&
\includegraphics[width=0.4\textwidth]{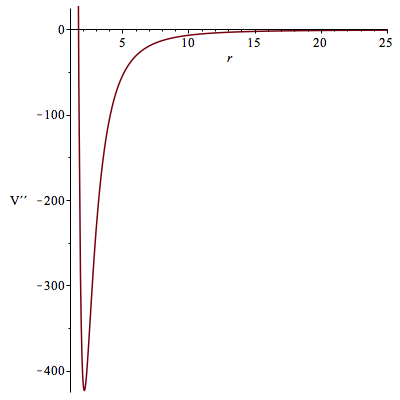}\\
 (a) & (b) \\
 \end{array}$$
\caption{ \label{fig:figureeffectpot}
  Plot  for the  variation of  $V''(r) \equiv \frac{d^2V}{ds^2}$  for  the two first members  of  the  family  of solutions (a)  $U_0$  and (b) $U_1$ for values of 
the parameters  given  by $\beta=0.25$,  $L=a=k_{\omega} = b_0=b_1=1$  and $E=50$.}
\end{figure}

For the sake of saving text space, we do not present here   the explicit  expression for  the derivative $\frac{d^2V}{ds^2}|_{r=r_0}$ and  the  corresponding  values  of the  
quantities $E$  and $L$.  However, in Fig.(\ref{fig:figureeffectpot}) we have  plotted  it for the  two  first  members of  the family  of   harmonic solutions $U_0$  and $U_1$ 
given by (\ref{eq:GeneralizedKuzmin}). The  plots  show  that  the  second  derivative  is  negative  for arbitrary  ranges  of $r$. So, the  orbits  of  the  test particles must 
be  stable in the halo. The  plots  also reveal  that the  orbits are unstable near  the center of  the  disk. We have also computed these derivatives for other values of the 
angular momentum $L$,  in all cases we have found a similar behavior. Moreover, a  simple but  long calculation shows that $\frac{d^2r}{ds^2} < 0$. Thus  the  particles are 
attracted towards the  center.

\section{Concluding remarks} \label{sec:conclude}
We used  the  formalism presented in   \cite{PhysRevD.87.044010}  to model  an  exact relativistic  rotating  disk surrounded by  a  magnetized  halo.  The  model  
was  obtained by  solving  the Einstein-Maxwell distributional field equations. In doing so, we introduced an auxiliary harmonic function that determines the functional dependence 
of the  metric components and the electromagnetic potential. Accordingly, we separated the  total energy-momentum tensor of the system  disk-halo. Additionally, we 
expressed the energy  momentum tensor of  the  halo  as a  sum of  two distributional contributions, one  due to the electromagnetic  part and the  other due to a  
material part. As we can see, due to  the spacetime  here considered is non-static (conformastationary),  the distributional  approach of the Einstein-Maxwell equations allows 
us to work with a strongly non-lineal partial  equations system.  We  considered,  for simplicity, the astrophysical consistent   case  in that there is  no electric  charge on the 
 halo. We obtained  that  the  charge density on  the  disk   is zero.

In order to analyze the physical content of the energy-momentum tensor of the halo and disk,  we projected each tensor,  in the canonical form, in the comoving frame 
defined  by the local observers tetrad. This analysis  allowed us  to  give  a  complete  dynamical description  of the  system in terms of two parameters (i.e $\beta$ and  
$k_\omega$) which  determine  the  matter  content  of  the sources.  Indeed, the parameter $\beta$ in the metric vanishes when it is equal to  the isotropic pressure and the 
anisotropic tensor on the material constituting the  disk. Similarly, when  the parameter $k_{\omega}$ is equal to zero the heat flux on the disk  and  halo 
vanishes, a feature of the static  systems. So, in this  paper we  presented,  for  first time, the complete analysis  of  the most  general energy-momentum tensor of  a   
disk-haloes system obtained   from an  exact conformastationary axially symmetric solutions  of  the  Einstein-Maxwell equations.

The expressions obtained here  are the  generalization of the obtained  for the conformastatic  disk-haloes  without isotropic pressure, stress tensor or heat flow presented in 
\cite{PhysRevD.87.044010}.  Moreover, when we take simultaneously $k_{\omega}= 0$ and $\beta =1$, we obtain its corresponding  electrized disk-haloes version. Furthermore,  our  
results are compatibles with the description of  the relativistic models of perfect fluid disks in a magnetic field presented  in \cite{garcia2014exact} and   the  halo  
presented in \cite{Chakraborty:2014paa}. Furthermore,  we have shown that the description of the  motion of charged  particles on the disk and is  in agreement with 
the  results of analysis of particles motion in the  magnetized disks discussed  in \cite{garcia2014exact}. 
In  accordance with  the  results presented  in \cite{Chakraborty:2014paa, nandi2009features}, we  also shown that  the  orbits  of  test  particles are  stable in the halo
for arbitrary  ranges  of $r$ and  unstable near  the center of  the  disk. It is also worth noticing that, one  can  fix the  values  of the 
parameters $\beta$, $k_{\omega}$ and $b_n$ as well  as the number of  members  of  the particular solutions presented  here  in order to have velocity increasing linearly with 
radius of  the  disk.

We  have considered  specific solutions  in which the gravitational and magnetic potential are completely determined  by  a ``generalization'' of  the Kuzmin-disk potential.   
Accordingly,  we  have generated  relativistic exact solutions  for magnetized  haloes surrounding rotating disks  from  a  
Newtonian gravitational  potential of a static axisymmetric distribution of  matter.  The  solution obtained is asymptotically  Minkowskian in general and  turns out to be free of  
singularities.

In short,  we   concluded that  we  have  presented   an exact general  relativistic  well-behaviored   rotating   disk  surrounded by a  well-behaviored  magnetized  
halo ``material".  In  our  description  we  do  not impose  restriction  on  the  kind of  ``material''  constituting  the system   disk-halo. Consequently, we  can 
speculate that the   halo  could be made of magnetized  dark matter.
This work   provides a solid footing to refine future studies of relativistic  disk-haloes systems.

\section*{Acknowledgement}
The author also wishes to acknowledge  useful discussions  with  C. S. Lopez-Monsalvo and H. Quevedo.

\appendix
\section{The local  observers}
We  write the  metric (\ref{eq:met1}) in the form
      \begin{eqnarray}
          \mathrm ds^2 = -   F( \mathrm dt  + \omega  \mathrm d\varphi)^2 +   F^{-\beta}  [\mathrm dr^2 + \mathrm  dz^2 + r^2{\mathrm d}\varphi^2  ], \label{eq:met2}
             \end{eqnarray}
where we have rewritten  $F=e^{2\phi}$.  In addition, we define the tetrad of the  local observers $h^{(\alpha)}_{\;\;\mu}$, in which the  metric  has  locally the  form of 
Minkowskian metric
       \begin{eqnarray}
           ds^2=\eta_{(\mu)(\nu)}\mathbf{h}^{( \mu)}\otimes\mathbf{h}^{ (\nu)},
          \end{eqnarray}
is  given  by
          \begin{subequations}\begin{eqnarray}
                {h}^{(t)}_{\quad \alpha}&=&F^{1/2}\{ 1,0,0, \omega \},\\
                {h}^{(r)}_{\quad\alpha}&=&F^{-\beta/2}\{ 0,1,0, 0 \},\\
                {h}^{(z)}_{\quad\alpha}&=&F^{-\beta/2}\{ 0,0,1,0 \},\\
                {h}^{(\varphi)}_{\quad\alpha} &=&F^{-\beta/2}\{ 0,0,0,r \}.\\
                \end{eqnarray}\label{eq:localobservator}\end{subequations}
The dual  tetrad reads
               \begin{subequations}\begin{eqnarray}
               {h}_{(t)}^{\quad \alpha}&=&F^{-1/2}\{ 1,0,0, 0 \},\\
               {h}_{(r)}^{\quad\alpha}&=&F^{\beta/2}\{ 0,1,0, 0 \},\\
               {h}_{(z)}^{\quad\alpha}&=&F^{\beta/2}\{ 0,0,1,0 \},\\
               {h}_{(\varphi)}^{\quad\alpha} &=&\frac{F^{\beta/2}}{r}\{ -\omega,0,0,1 \}.\\
                     \end{eqnarray}\label{eq:duallocalobservator}\end{subequations}

The  circular velocity of  the system  disk-halo can be  modelled  by a fluid space-time whose circular velocity
$V^{\alpha}$ can be written in terms of two Killing  vectors $t^{\alpha}$ and $\varphi^{\alpha}$,
            \begin{eqnarray}
                V^{\alpha}=V^t(t^{\alpha} + \Omega\varphi^{\alpha})\label{eq:velocity},
                \end{eqnarray}
where
            \begin{eqnarray}
               \Omega \equiv\frac{u^{\varphi}}{u^t}=\frac{d\varphi}{dt}
                  \end{eqnarray}
is  the  angular velocity  of  the  fluid  as  seen  by  an  observer at rest at infinity. The  velocity satisfy the  normalization $V_{\alpha}V^{\alpha}=-1$. 
Accordingly, for  the metric (\ref{eq:met2} ) we  have
             \begin{eqnarray}
                (V^{t})^2=\frac{1}{-t^{\alpha}t_{\alpha} - 2\Omega t^{\alpha}{\varphi}_{\alpha} - \Omega{\varphi}^{\alpha}{\varphi}_{\alpha}},\label{eq:cuadrivelocity}
               \end{eqnarray}
with
             \begin{subequations}\begin{eqnarray}
                t^{\alpha}t_{\alpha}&=&g_{tt}=- F\\ 
                t^{\alpha}{\varphi}_{\alpha}&=&g_{t\varphi}=-\omega F\\
                {\varphi}^{\alpha}{\varphi}_{\alpha}&=&g_{\varphi\varphi}=r^2F^{-\beta}(1 - F^{1+\beta}\frac{\omega^2}{r^2}),
                     \end{eqnarray}\end{subequations}
consequently we  write  the  velocity  as
            \begin{eqnarray} 
               V^t=\frac{1}{F^{1/2}(1 +\omega\Omega)\sqrt{1 - V_{_{LOC}}^2}},
                 \end{eqnarray}
where
            \begin{eqnarray}
                 V_{_{LOC}}\equiv \frac{r\Omega}{F^{{(1+\beta)}/{2}}(1 +\omega\Omega)},\label{eq:localvelocity}
                 \end{eqnarray}
 is the  velocity as measured  by  the local observers.

 
 %

\end{document}